# Soliton dynamics in an extended nonlinear Schrödinger equation with a spatial counterpart of the stimulated Raman scattering

**E.M. Gromov**[1a] **and B.A. Malomed**[b]

[a]*National Research University Higher School of Economics, 25/12 Bolshaja Pecherskaja Ulitsa, Nizhny Novgorod 603155, Russia*

[b]*Department of Physical Electronics, Faculty of Engineering, Tel Aviv University, Tel Aviv, 69978, Israel*

Dynamics of solitons is considered in the framework of the extended nonlinear Schrödinger equation (NLSE), which is derived from a system of Zakharov's type for the interaction between high- and low-frequency (HF and LF) waves, in which the LF field is subject to diffusive damping. The model may apply to the propagation of HF waves in plasmas. The resulting NLSE includes a *pseudo-stimulated-Raman-scattering* (PSRS) term, i.e., a spatial-domain counterpart of the SRS term which is well known as an ingredient of the temporal-domain NLSE in optics. Also included is inhomogeneity of the spatial second-order diffraction (SOD). It is shown that the wavenumber downshift of solitons, caused by the PSRS, may be compensated by an upshift provided by the SOD whose coefficient is a linear function of the coordinate. An analytical solution for solitons is obtained in an approximate form. Analytical and numerical results agree well, including the predicted balance between the PSRS and the linearly inhomogeneous SOD.

## 1. Introduction
The great interest to the dynamics of solitons is motivated by their ability to travel large distances, preserving a permanent shape. Soliton solutions are generated by various nonlinear models dealing with the propagation of intense wave fields in dispersive media: optical pulses in fibers, electromagnetic waves in plasma, surface waves on deep water, etc. (Agrawal 2001; Kivshar & Agrawal 2003; Dauxois & Peyrard 2006; Dickey 2005; Infeld & Rowlands 2000; Malomed 2006; Yang 2001).

Dynamics of long high-frequency (HF) wave packets is described by the second-order nonlinear dispersive wave theory. The fundamental equation of the theory is the nonlinear Schrödinger equation (NLSE) (Zakharov & Shabat 1972; Hasegava & Tappert 1973), which includes the second-order dispersion (SOD) and cubic nonlinearity (self-phase modulation). Soliton solutions in this case arise as a result of the balance between the dispersive stretch and nonlinear compression of the wave packet.

The dynamics of short HF wave packets is described by the third-order nonlinear dispersive wave theory (Agrawal 2001), which takes into account the nonlinear dispersion (self-steeping) (Oliviera & Moura 1998), stimulated Raman scattering (SRS) in optical media (Mitschke & Mollenauer 1986; Gordon 1986; Kodama 1985), and third-order dispersion (TOD). Accordingly, the basic equation of the theory is the third-order NLSE (Kodama 1985; Kodama & Hasegava 1987; Zaspel 1999; Karpman 2004; Hong & Lu 2009). Soliton solutions in the framework of the third-order NLSE without the SRS term were investigated in Refs. (Gromov & Talanov 1996, 2000; Gromov, Piskunova & Tyutin 1999; Obregon & Stepanyants 1998; Scalora *at al* 2005; Wen *at al* 2006; Marklund, Shukla & Stenflo 2006; Tsitsas *at al* 2009). Such solitons are supported by the balance between the TOD and nonlinear dispersion. In (Kivshar 1990; Kivshar & Malomed 1993) stationary kink modes were found as solutions of the third-order NLSE without the TOD term. This solution exists due to the balance between the nonlinear dispersion and SRS. For localized nonlinear wave packets (solitons), the SRS gives rise to the downshift of the soliton spectrum (Kodama 1985; Gordon 1986; Mitschke & Mollenauer 1986) and eventually to destabilization of the solitons. The use of the balance between the SRS and the slope of the gain for the stabilization of solitons in long telecom links was proposed by Malomed & Tasgal (1998). The compensation of the SRS by emission of linear radiation fields from the soliton's core was considered in by Biancalama, Skrybin & Yulin (2004). The compensation of the SRS in inhomogeneous media was considered in other settings too: periodic SOD (Essiambre & Agraval 1997a, 1997b), sliding zero-dispersion point (Andrianov, Muraviev, Kim & Sysoliatin 2007), and dispersion-decreasing fibers (Chernikov, Dianov, Richardson & Payne 1993).

---
[1] Email address for correspondence: egromov@hse.ru

In this work the dynamics of HF wave packets is considered in dispersive nonlinear media, taking into account the interaction of the HF field with low-frequency (LF) waves, which are subject to diffusive damping. In the third-order approximation of the dispersion-wave theory, the original Zakharov-type system of equations for the HF and LF fields (the LF equation includes the diffusion/viscosity term), is reduced to an extended NLSE, which features a spatial counterpart of the SRS term, that we call a *pseudo-SRS* (PSRS) one. The model may be realized for the propagation of nonlinear waves in plasmas. The PSRS leads to the self–wavenumber downshift, similar to what is well known in the temporal domain (Agrawal 2001; Gordon 1986; Kodama 1985; Mitschke & Mollenauer 1986). On the other hand, the SOD term with a spatially decreasing coefficient leads to an increase of the soliton's wavenumber. The balance between the PSRS and the decreasing SOD leads to stabilization of the soliton's wavenumber spectrum. An analytical soliton solution is found in an approximate form.

The NPSE is derived in Section 2 and analytical results are presented in Section 3. Numerical findings, and their comparison to the analytical predictions, are summarized in Section 4. The paper is concluded by Section 5.

## 2. The basic equation and integrals relations

We consider the evolution of slowly varying envelope $U(\xi,t)$ of the intense HF wave field in the nonlinear medium with inhomogeneous SOD, taking into account the interaction with LF variations of the medium's parameter $n(\xi,t)$ (such as the refractive index in optics), which suffers the action of effective diffusion. The unidirectional propagation of the fields along coordinate $\xi$ is described by the system of the Zakharov's type (Zakharov 1971, 1974):

$$2i\frac{\partial U}{\partial t} + \frac{\partial}{\partial \xi}\left(q(\xi)\frac{\partial U}{\partial \xi}\right) - nU = 0, \quad (1)$$

$$\frac{\partial n}{\partial t} + \frac{\partial n}{\partial \xi} - \mu\frac{\partial^2 n}{\partial \xi^2} = -\frac{\partial |U|^2}{\partial \xi}, \quad (2)$$

where $\mu$ is the diffusion coefficient. In particular, this system may describe intense electromagnetic or Langmuir waves in plasmas, taking into account the scattering on ion-acoustic waves, which are subject to the viscous damping. The second-order approximation of the dispersion-wave theory corresponds to replacing Eq. (2) by adiabatic relation $n = -|U|^2$, hence envelope $U$ of the HF wave packet obeys the NSLE: $2i\partial U/\partial t + \partial(q(\xi)\partial U/\partial \xi)/\partial \xi + 2\alpha U|U|^2 = 0$, where $\alpha = 1/2$.

In the third-order approximation of the theory (for short HF wave packets, with $k\Delta << \mu$, where $k$ and $\Delta$ are the spatial extension and characteristic wave number of the wave packet), Eq. (2) may approximated by the nonlinear response of the medium, $n = -|U|^2 - \mu \partial(|U|^2)/\partial \xi$, which leads to the following extended NLSE for the HF amplitude:

$$2i\frac{\partial U}{\partial t} + \frac{\partial}{\partial \xi}\left(q(\xi)\frac{\partial U}{\partial \xi}\right) + 2\alpha U|U|^2 + \mu U \frac{\partial(|U|^2)}{\partial \xi} = 0. \qquad (3)$$

The last term in Eq. (3) represents the above-mentioned PSRS effect in the spatial domain.

Equation (3) with zero boundary conditions at infinity, $U|_{\xi \to \pm\infty} \to 0$, gives rise to the following integral relations for field moments, which will be used below:

– the conservation of the wave action, $N$ (a well-known property of the SRS term):

$$\frac{dN}{dt} \equiv \frac{d}{dt}\int_{-\infty}^{+\infty}|U|^2 d\xi = 0, \qquad (4)$$

– the rate of change of the wave-field momentum:

$$2\frac{d}{dt}\int_{-\infty}^{+\infty} K|U|^2 d\xi = -\mu \int_{-\infty}^{+\infty}\left(\frac{\partial(|U|^2)}{\partial \xi}\right)^2 d\xi - \int_{-\infty}^{+\infty}\frac{\partial q}{\partial \xi}\left|\frac{\partial U}{\partial \xi}\right|^2 d\xi, \qquad (5)$$

– the rate of change of the integrated squared absolute value of the gradient of the wave field:

$$\frac{d}{dt}\int_{-\infty}^{+\infty}\left|\frac{\partial U}{\partial \xi}\right|^2 d\xi = -\mu \int_{-\infty}^{+\infty} K\left(\frac{\partial |U|^2}{\partial \xi}\right)^2 d\xi + \alpha \int_{-\infty}^{+\infty} K\frac{\partial |U|^4}{\partial \xi} d\xi - \int_{-\infty}^{+\infty}\frac{\partial q}{\partial \xi} K\left(3\left|\frac{\partial U}{\partial \xi}\right|^2 - \frac{1}{2}\frac{\partial^2 |U|^2}{\partial \xi^2} - 2K^2|U|^2\right)d\xi, \quad (6)$$

– the rate of change of the squared gradient of the wave-field intensity:

$$\frac{d}{dt}\int_{-\infty}^{\infty}\left(\frac{\partial(|U|^2)}{\partial \xi}\right)^2 d\xi = -2\int_{-\infty}^{+\infty}\frac{\partial^2(|U|^2)}{\partial \xi^2}\frac{\partial(qK|U|^2)}{\partial \xi} d\xi, \qquad (7)$$

– the equation of motion for the center-of-mass coordinate, $\bar{\xi} \equiv N^{-1}\int_{-\infty}^{+\infty}\xi|U|^2 d\xi$:

$$\frac{d}{dt}\int_{-\infty}^{+\infty}\xi|U|^2 d\xi = \int_{-\infty}^{+\infty} qK|U|^2 d\xi, \qquad (8)$$

where the complex field is represented as $U \equiv |U|\exp(i\phi)$, and $K \equiv \partial \phi/\partial \xi$ is the local wavenumber.

## 3. Analytical results

For analytical consideration of the wave-packet dynamics, we assume that the scales of the inhomogeneity of both the SOD term and local wavenumber $K$ are much larger than the size of the wave-packet envelope, hence the spatial variation of the wavenumber may be locally approximated by the linear function of the coordinate, $K(\xi,t) \approx K(\bar{\xi},t) + (\partial K/\partial \xi)_{\bar{\xi}}(\xi - \bar{\xi})$. Then we obtain from the imaginary part of Eq. (3) under condition $(\partial |U|/\partial \xi)_{\bar{\xi}} = 0$ (which means that the peak of the soliton's amplitude is located at its center):

$$\left(\frac{\partial K}{\partial \xi}\right)_{\bar{\xi}} = -\left(\frac{2}{q|U|}\frac{\partial |U|}{\partial t} + \frac{1}{q}\frac{dq}{d\xi}K\right)_{\bar{\xi}}. \qquad (9)$$

Further, replacing $K(\xi,t)$ for soliton-like wave packets by $K(\bar{\xi},t) \equiv k(t)$, the system of Eqs. (4)-(7) can be cast in the form of evolution equations for parameters of the wave packet:

$$2\frac{dk}{dt} = -\mu\frac{L_0}{N}l - q'(\bar{\xi})z, \qquad (10)$$

$$\frac{dz}{dt} = -\mu\frac{L_0}{N}kl - 3kq'(\bar{\xi})z + 2k^3 q'(\bar{\xi}), \qquad (11)$$

$$\frac{dl}{dt} = -3kq'(\bar{\xi})l. \qquad (12)$$

Here $q'(\bar{\xi}) = (dq/d\xi)_{\bar{\xi}}$ is the gradient of the SOD coefficient at the center of the packet, $l = L/L_0$, $z = Z/N$, and $Z \equiv \int_{-\infty}^{+\infty} |\partial U/\partial\xi|^2 d\xi$, $L \equiv \int_{-\infty}^{+\infty} (\partial|U|^2/\partial\xi)^2 d\xi$, along with the conserved wave action $N$, are integral characteristics of the wave packet, $Z_0 = Z(0)$, $L_0 = L(0)$ being their initial values. An equilibrium state (fixed point) of Eqs. (10)-(12) corresponds to conditions

$$k = 0, \ \mu L_0 = -q'(\bar{\xi}_0)Z_0. \qquad (13)$$

To analyze the dynamics of the wave packet with non-equilibrium parameters, we assume that the SOD coefficient is a linear function of the coordinate,

$$q(\xi) = q_0 + q'\xi. \qquad (14)$$

Then, substitutions $\tau \equiv -tq'$ and $p \equiv \mu L_0/(-q'N_0)$ reduce Eqs. (10)-(12) to

$$2\frac{dk}{d\tau} = z - pl, \quad \frac{dz}{d\tau} = (3z - pl - 2k^2)k, \quad \frac{dl}{d\tau} = 3kl. \qquad (15)$$

The first integral of Eqs. (15) is

$$2k^2/z_0 + (2 - \lambda - 4k_0^2/z_0)\sqrt[3]{l} - 2(1 - k_0^2/z_0)\sqrt[3]{l^2} + \lambda l = 0, \qquad (16)$$

where $k_0 = k(0)$, $\lambda \equiv p/z_0$, $z_0 \equiv Z_0/N$. In Fig. 1, this relation between variables $k/\sqrt{z_0}$ and $l$ is plotted for $k_0 = 0$ and different values of $\lambda$.

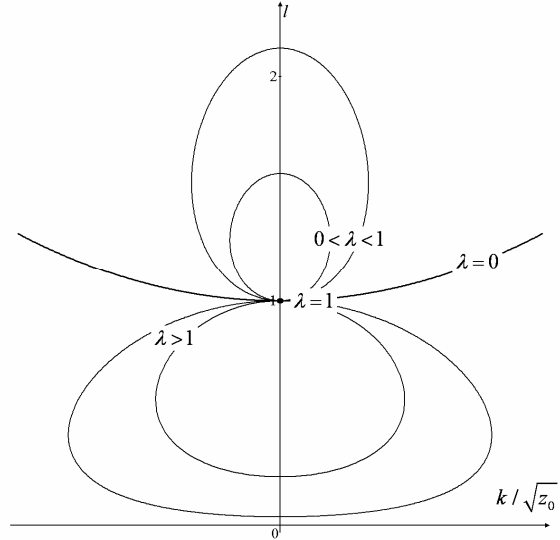

Figure 1. Relation (16) for the soliton variables in the plane of $(k/\sqrt{z_0},\ l)$ for $k_0 = 0$ and different values of $\lambda$.

We now look for stationary solutions to Eq. (3), where the SOD with spatial profile (14) is adopted, in the form of a stationary wave profile, $U(\xi,t) = \psi(\xi)\exp(i\Omega t)$:

$$(q_0 + q'\xi)\frac{d^2\psi}{d\xi^2} + q'\frac{d\psi}{d\xi} + 2\alpha\psi^3 - 2\Omega\psi + \mu\psi\frac{d(\psi^2)}{d\xi} = 0. \qquad (17)$$

Next, with regard to the underlying assumption that the soliton's width is much smaller than the scale of the spatial inhomogeneity for the SOD than for the packet's envelope, a solution to Eq. (17) is found in the form of $\psi = \psi_0 + \psi_1$, where $\psi_1$ is a small correction to $\psi_0$. In this approximation, we obtain

$$q_0 \frac{d^2\psi_0}{d\xi^2} + 2\alpha\psi_0^3 - 2\Omega\psi_0 = 0, \tag{18}$$

$$q_0 \frac{d^2\psi_1}{d\xi^2} + \left(6\alpha\psi_0^2 - 2\Omega\right)\psi_1 = -q'\frac{d^2\psi_0}{d\xi^2}\xi - \frac{2}{3}\mu\frac{d(\psi_0^3)}{d\xi} - q'\frac{d\psi_0}{d\xi}. \tag{19}$$

Equation (18) gives rise to the classical soliton solution, $\psi_0 = A_0 \operatorname{sech}(\xi/\Delta)$, where $\Delta \equiv \sqrt{q_0/\alpha}/A_0$ and $\Omega \equiv \alpha A_0^2/2$. Then substitutions $\eta = \xi/\Delta$ and $\Psi = \psi_1 q_0/(-A_0 q'_\eta)$ cast Eq. (19) in the form of

$$\frac{d^2\Psi}{d\eta^2} + \left(\frac{6}{\cosh^2\eta} - 1\right)\Psi = 2\frac{\eta}{\cosh^3\eta} - \frac{\eta}{\cosh\eta} + \frac{5}{4}\frac{\mu}{\mu_*}\frac{\sinh\eta}{\cosh^4\eta} + \frac{\sinh\eta}{\cosh^2\eta}, \tag{20}$$

where the equilibrium value of the PSRS coefficient is

$$\mu_* \equiv -5q'_\eta \sqrt{\alpha/q_0}/(8A_0). \tag{21}$$

Under condition $\Psi(0)=0$, an exact solution to Eq. (20) can be found,

$$\Psi(\eta) = \left(\Psi'(0)\eta - \frac{\eta^2}{4}\tanh\eta + \frac{\mu}{4\mu_*}(\tanh\eta)\ln(\cosh\eta)\right)\operatorname{sech}\eta + \frac{1}{12}\left(\frac{\mu}{\mu_*} - 1\right)(\tanh^2\eta)\sinh\eta, \tag{22}$$

cf. a similar solution reported by Blit & Malomed (2012). At $\mu = \mu_*$, it satisfies boundary conditions $\Psi(\eta \to \pm\infty) \to 0$. This spatially antisymmetric solution, which is displayed in Fig. 2 for $\mu = \mu_*$ and different values of $\Psi'(0)$, exists due to the balance between the PSRS term and linearly decreasing SOD. At $\mu \neq \mu_*$, solution (22) diverges at the spatial infinity.

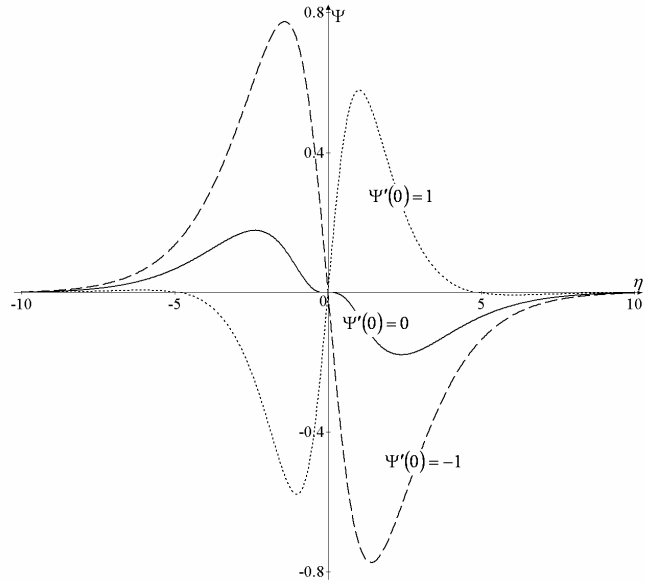

Figure 2. Solution (22) for $\Psi(0)=0$, $\mu = \mu_*$, and different values of $\Psi'(0)$.

## 4. Numerical results

We numerically solved Eq. (3) with units fixed by setting $\alpha \equiv 1$, different values of the PSRS coefficient $\mu$, linear SOD profile $q(\xi) = 1 - \xi/10$ (cf. Eq. (14)), and initial conditions corresponding to the soliton, $U(\xi, t=0) = \operatorname{sech}\xi$. In this case, Eq. (22) predicts the equilibrium value of the PSRS coefficient, $\mu_* = 1/16$, for this initial pulse. As shown in Fig. 3, in direct simulations of Eq. (3) with $\mu = 1/16$ the input evolves into a stationary localized pulse with zero wavenumber, which is close to the above analytical solution, see Eq. (22), with $q_0 = \alpha = A_0 = 1$, $q' = -1/10$, $\mu = \mu_*$:

$$|U| = \left(1 + \frac{1}{40}\left((\tanh \xi)\ln(\cosh \xi) - \xi^2 \tanh \xi\right)\right)\operatorname{sech}\xi. \qquad (23)$$

In Fig. 3, the profile of the soliton solution (23) is shown by the dotted curve.

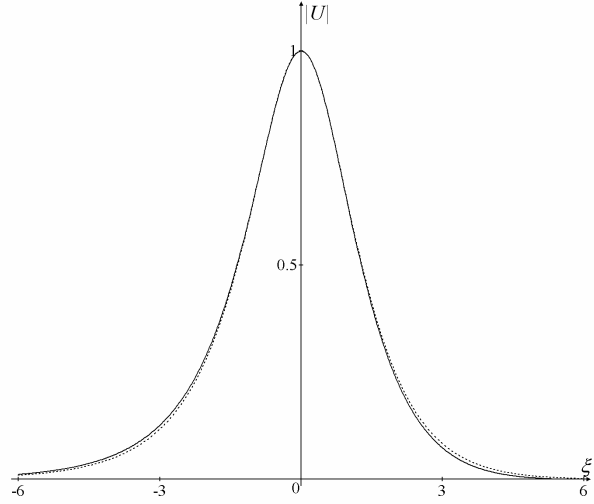

Fig. 3. The solid curve is the result of the numerical solution of Eq. (3) for the soliton's envelope, $|U(\xi)|$, obtained in this stationary form in the time interval $10 < t \leq 300$, for $q(\xi) = 1 - \xi/10$ and $\mu = 1/16$. The dotted curve is the analytical solution (23).

At values of the PSRS coefficient different from $\mu_*$, the simulations produce nonstationary solitons, with variable amplitude and wavenumber, see an example for $\mu = 5/128 \equiv (5/8)\mu_*$ in Figs. 4,5

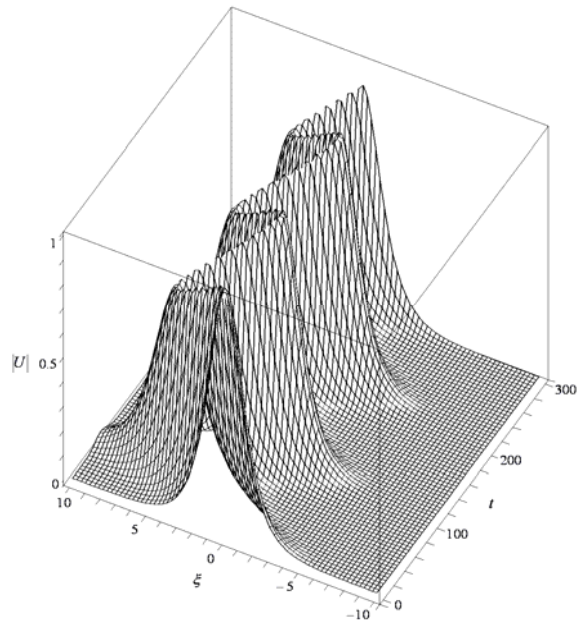

Figure 4. The numerically simulation evolution of the wave-packet envelope versus $\xi, t$ for $\mu = 5/128$.

In Fig. 5, numerical results produced, as functions of time, by the simulations for the local wavenumber at the maximum of the wave-packet's shape, are compared with the analytical counterparts obtained from Eq. (10) for different values of $\mu$. Close agreement between the analytical and numerical results is observed by the figure, both for $\mu = \mu_*$, when both the numerically and analytically found wavenumbers remain equal to zero, and for nonstationary pulses at $\mu \neq \mu_*$. A similar picture is observed at other values of the parameters.

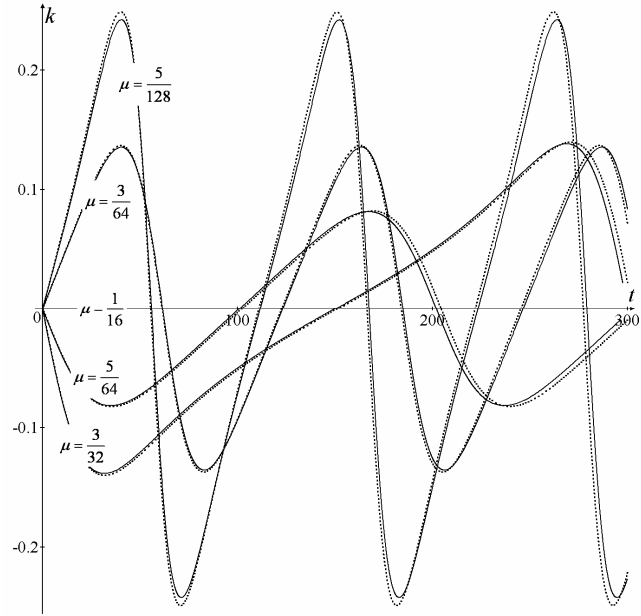

Figure 5. Numerical and analytical results (solid and dashed curves) for the local wavenumber at the point of the maximum of the wave-packet envelope versus $t$ for the SOD profile $q(\xi) = 1 - \xi/10$ and different values of the PSRS coefficient $\mu$.

## 5. Conclusion

We have proposed a model which realizes the extended NLSE with the spatial-domain counterpart of the SRS term (the PSRS, i.e., pseudo-SRS one). The equation is derived from the system of the Zakharov's type for electromagnetic and Langmuir waves in plasmas, in which the LF field is subject to the diffusive damping. We have studied the soliton dynamics is the framework of the extended NLSE, which also includes the smooth spatial variation of the SOD (second-order dispersion) coefficient. The analytical predictions were produced by integral relations for the field moments, and numerical results were generated by systematic simulations of the pulse evolution in the framework of the extended NLSE. Stable stationary solitons are maintained by the balance between the self-wavenumber downshift, caused by the PSRS, and the upshift induced by the linearly decreasing SOD. The analytical solutions are found to be in close agreement with their numerical counterparts.

In this work the soliton dynamics was considered in the model neglecting the nonlinear dispersion and third-order linear dispersion. These effects will be considered elsewhere. It may also be interesting to study interactions between solitons in the present model.